\documentclass[aps,twocolumn,showpacs,preprintnumbers,amsmath,amssymb,groupedaddress,superscriptaddress]{revtex4}
\usepackage{graphicx}
\usepackage{dcolumn}
\usepackage{bm}
\usepackage{lineno}
\usepackage{xspace} % dynamic space in definitions

\newcommand{\sacs}{{$\left<\sigma_{\rm A} v\right>$}}
\newcommand{\noff}{N^{ \mbox{\rm \tiny{off}} }_{\rm \mbox{\tiny{BG}}}}
\newcommand{\non} {N^{ \mbox{\rm \tiny{on}} }_{\rm \mbox{\tiny{BG}}}}

\begin{document}
\preprint{APS/123-QED}

\title{Search for Neutrinos from Annihilating Dark Matter in the Direction of the Galactic Center with the 40-String IceCube Neutrino Observatory}

\affiliation{III. Physikalisches Institut, RWTH Aachen University, D-52056 Aachen, Germany}
\affiliation{School of Chemistry \& Physics, University of Adelaide, Adelaide SA, 5005 Australia}
\affiliation{Dept.~of Physics and Astronomy, University of Alaska Anchorage, 3211 Providence Dr., Anchorage, AK 99508, USA}
\affiliation{CTSPS, Clark-Atlanta University, Atlanta, GA 30314, USA}
\affiliation{School of Physics and Center for Relativistic Astrophysics, Georgia Institute of Technology, Atlanta, GA 30332, USA}
\affiliation{Dept.~of Physics, Southern University, Baton Rouge, LA 70813, USA}
\affiliation{Dept.~of Physics, University of California, Berkeley, CA 94720, USA}
\affiliation{Lawrence Berkeley National Laboratory, Berkeley, CA 94720, USA}
\affiliation{Institut f\"ur Physik, Humboldt-Universit\"at zu Berlin, D-12489 Berlin, Germany}
\affiliation{Fakult\"at f\"ur Physik \& Astronomie, Ruhr-Universit\"at Bochum, D-44780 Bochum, Germany}
\affiliation{Physikalisches Institut, Universit\"at Bonn, Nussallee 12, D-53115 Bonn, Germany}
\affiliation{Universit\'e Libre de Bruxelles, Science Faculty CP230, B-1050 Brussels, Belgium}
\affiliation{Vrije Universiteit Brussel, Dienst ELEM, B-1050 Brussels, Belgium}
\affiliation{Dept.~of Physics, Chiba University, Chiba 263-8522, Japan}
\affiliation{Dept.~of Physics and Astronomy, University of Canterbury, Private Bag 4800, Christchurch, New Zealand}
\affiliation{Dept.~of Physics, University of Maryland, College Park, MD 20742, USA}
\affiliation{Dept.~of Physics and Center for Cosmology and Astro-Particle Physics, Ohio State University, Columbus, OH 43210, USA}
\affiliation{Dept.~of Astronomy, Ohio State University, Columbus, OH 43210, USA}
\affiliation{Dept.~of Physics, TU Dortmund University, D-44221 Dortmund, Germany}
\affiliation{Dept.~of Physics, University of Alberta, Edmonton, Alberta, Canada T6G 2G7}
\affiliation{D\'epartement de physique nucl\'eaire et corpusculaire, Universit\'e de Gen\`eve, CH-1211 Gen\`eve, Switzerland}
\affiliation{Dept.~of Physics and Astronomy, University of Gent, B-9000 Gent, Belgium}
\affiliation{Dept.~of Physics and Astronomy, University of California, Irvine, CA 92697, USA}
\affiliation{Laboratory for High Energy Physics, \'Ecole Polytechnique F\'ed\'erale, CH-1015 Lausanne, Switzerland}
\affiliation{Dept.~of Physics and Astronomy, University of Kansas, Lawrence, KS 66045, USA}
\affiliation{Dept.~of Astronomy, University of Wisconsin, Madison, WI 53706, USA}
\affiliation{Dept.~of Physics and Wisconsin IceCube Particle Astrophysics Center, University of Wisconsin, Madison, WI 53706, USA}
\affiliation{Institute of Physics, University of Mainz, Staudinger Weg 7, D-55099 Mainz, Germany}
\affiliation{Universit\'e de Mons, 7000 Mons, Belgium}
\affiliation{T.U. Munich, D-85748 Garching, Germany}
\affiliation{Bartol Research Institute and Department of Physics and Astronomy, University of Delaware, Newark, DE 19716, USA}
\affiliation{Dept.~of Physics, University of Oxford, 1 Keble Road, Oxford OX1 3NP, UK}
\affiliation{Dept.~of Physics, University of Wisconsin, River Falls, WI 54022, USA}
\affiliation{Oskar Klein Centre and Dept.~of Physics, Stockholm University, SE-10691 Stockholm, Sweden}
\affiliation{Department of Physics and Astronomy, Stony Brook University, Stony Brook, NY 11794-3800, USA}
\affiliation{Dept.~of Physics and Astronomy, University of Alabama, Tuscaloosa, AL 35487, USA}
\affiliation{Dept.~of Astronomy and Astrophysics, Pennsylvania State University, University Park, PA 16802, USA}
\affiliation{Dept.~of Physics, Pennsylvania State University, University Park, PA 16802, USA}
\affiliation{Dept.~of Physics and Astronomy, Uppsala University, Box 516, S-75120 Uppsala, Sweden}
\affiliation{Dept.~of Physics, University of Wuppertal, D-42119 Wuppertal, Germany}
\affiliation{DESY, D-15735 Zeuthen, Germany}

\author{R.~Abbasi}
\affiliation{Dept.~of Physics and Wisconsin IceCube Particle Astrophysics Center, University of Wisconsin, Madison, WI 53706, USA}
\author{Y.~Abdou}
\affiliation{Dept.~of Physics and Astronomy, University of Gent, B-9000 Gent, Belgium}
\author{M.~Ackermann}
\affiliation{DESY, D-15735 Zeuthen, Germany}
\author{J.~Adams}
\affiliation{Dept.~of Physics and Astronomy, University of Canterbury, Private Bag 4800, Christchurch, New Zealand}
\author{J.~A.~Aguilar}
\affiliation{D\'epartement de physique nucl\'eaire et corpusculaire, Universit\'e de Gen\`eve, CH-1211 Gen\`eve, Switzerland}
\author{M.~Ahlers}
\affiliation{Dept.~of Physics and Wisconsin IceCube Particle Astrophysics Center, University of Wisconsin, Madison, WI 53706, USA}
\author{D.~Altmann}
\affiliation{Institut f\"ur Physik, Humboldt-Universit\"at zu Berlin, D-12489 Berlin, Germany}
\author{K.~Andeen}
\affiliation{Dept.~of Physics and Wisconsin IceCube Particle Astrophysics Center, University of Wisconsin, Madison, WI 53706, USA}
\author{J.~Auffenberg}
\affiliation{Dept.~of Physics and Wisconsin IceCube Particle Astrophysics Center, University of Wisconsin, Madison, WI 53706, USA}
\author{X.~Bai}
\thanks{Physics Department, South Dakota School of Mines and Technology, Rapid City, SD 57701, USA}
\affiliation{Bartol Research Institute and Department of Physics and Astronomy, University of Delaware, Newark, DE 19716, USA}
% ...

\begin{abstract}
A search for muon neutrinos from dark matter annihilations in the Galactic Center region has
been performed with the 40-string configuration of the IceCube Neutrino Observatory
using data collected in 367 days of live-time starting in April 2008.
The observed fluxes were consistent with the atmospheric background expectations.
Upper limits on the self-annihilation cross-section are obtained for dark matter particle masses ranging
from 100\,GeV to 10\,TeV. In the case of decaying dark matter, lower limits on the lifetime
have been determined for masses between 200\,GeV and 20\,TeV.
\end{abstract}

\maketitle
%\setpagewiselinenumbers
%\linenumbers

\subsection{\label{sec:Intro} Introduction}

\par
There is well-established observational evidence for the existence of a
non-baryonic cold dark matter component in the universe.
Although the exact nature of dark matter remains unknown, many theories beyond the
Standard Model of particle physics accommodate stable or extremely long-lived particles as
potential dark matter candidates~\cite{Bertone:2004pz}. A favored hypothesis is WIMPs (Weakly Interacting
Massive Particles), which are predicted to have masses from $\mathcal{O}(10)\,\mathrm{GeV}$
up to several TeV, with a theoretical upper limit of 340~TeV based on unitarity requirements~\cite{Griest:1989wd}.
Particles that have properties of a WIMP naturally arise from many theories, which were not
initially designed for this purpose.
For example, the neutralino in supersymmetric models or the lightest Kaluza-Klein
particle in theories with extra dimensions are natural WIMP candidates.

WIMPs could self-annihilate if they are Majorana particles, or decay, resulting
in stable particles like gamma-rays, electrons and neutrinos. A measurable
flux of such particles would be strong indirect evidence for the existence of dark matter.
Regions with suspected high density of dark matter are prominent targets for these indirect searches
since a large dark matter density corresponds to a higher flux of detectable particles.
Since galaxies are expected to be embedded in a halo of dark matter, the Galactic halo
and Galactic Center are examples of such expected high density regions.

Self-annihilation of dark matter particles in the Galaxy is of particular interest since
a WIMP search via the product particles is sensitive to the velocity-averaged
self-annihilation cross-section \sacs~\cite{Yuksel:2007ac}. In this sense, such
searches are complementary to direct searches or indirect searches from the Sun. 
Like the direct searches, the latter probes the WIMP-nucleon scattering cross-section,
since dark matter would be captured gravitationally due to repetitive elastic scattering. 
While the analysis presented in this paper is based on a search for an excess neutrino
flux from the direction of the Galactic Center, IceCube also published results from a search
for dark matter in the Galactic halo~\cite{Abbasi:2011eq}, as well as the Sun~\cite{PhysRevD.85.042002}.

A search for neutrino signals from dark matter annihilations has inherent advantages over
searches using photons or charged cosmic rays since neutrinos may escape the production
point and arrive at the detector unimpeded by absorption by matter or deflection by the
magnetic field in the central region of the Galaxy.\par

\subsection{\label{sec:simulations} Signal from Annihilating or Decaying Dark Matter}

A description of the dark matter density in the Milky Way is essential for a
calculation of the signal flux. There are various models that attempt to describe
the distribution profile of dark matter in Milky Way-size galaxies. These models
describe a dark matter halo with a spherically symmetric matter density,
$\rho_{\mbox{\tiny DM}}(r)$, that peaks at the center of the galaxy and decreases
with distance.
Profiles based on N-body simulations~\cite{Diemand:2007qr} of cold dark matter tend to 
show a divergent density in the central region. On the other hand, profiles that
rely on observations of low surface brightness galaxies dominated by dark matter
imply a flat distribution in the central region~\cite{Blok:2002tr, deBlok:2009sp}.
Despite large differences in the description of the central region, the models show a
similar density behavior at large distances from the Galactic Center (\mbox{$\gtrsim 8.5\,\mathrm{kpc}$}).

\begin{figure}[t]
  \centering
  \includegraphics[width=0.5\textwidth]{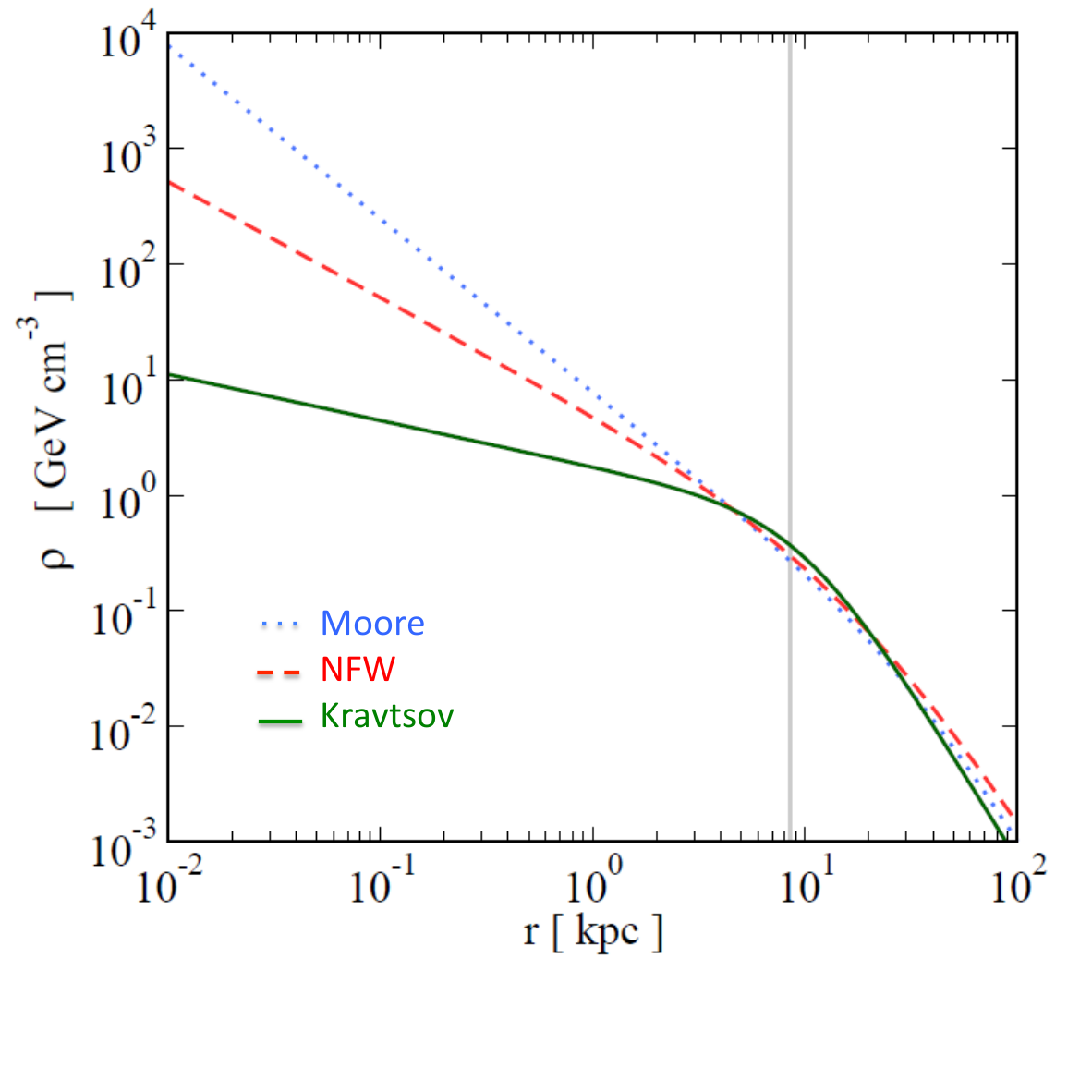}
  \caption{Dark matter density as a function of the distance to the Galactic Center for
  	three halo profiles~\cite{Yuksel:2007ac}. From top to bottom, the Moore (blue-dotted),
  	NFW (red-dashed) and Kravtsov (green-solid) profiles are shown. The vertical line marks
  	the orbital radius of the solar system.}
  \label{fig:HaloModels}
\end{figure}

Frequently considered models are the Navarro-Frenk-White (NFW) model~\cite{Navarro:1995iw},
the Moore et al. model~\cite{Moore:1999nt} and the Kravtsov et al. model~\cite{kravtsov:1997dp}.
We use the NFW profile as a benchmark for the analysis presented in this paper, with
the very cuspy Moore profile and the Kravtosv profile, which is close to isothermal
or cored profiles, as extreme cases. These models are compared in Fig.~\ref{fig:HaloModels}.

The NFW profile is given by 
\begin{equation}
\rho_{\mbox{\tiny DM}} (r) = {\rho_0 \over {r \over r_s} \left ( 1+{r \over r_s} \right )^2} ~ ,
\end{equation}
with $r_s=20 \,\textrm{kpc}$~\cite{Navarro:1995iw}. %Yuksel:2007ac, 
The normalization parameter, $\rho_0$, is chosen so that the dark matter density at
the orbital distance of the solar system ($R_{\mbox{\tiny SC}} = 8.5\,\mathrm{kpc}$)
is the assumed local density \mbox{$\rho_{\mbox{\tiny DM}}(R_{\mbox{\tiny SC}} ) = 
0.3\, \mathrm{GeV cm^{-3}}$}~\cite{Yuksel:2007ac}. The choice of parameters is
consistent with a previous analysis~\cite{Abbasi:2011eq} of signals from the outer
halo based on data from IceCube in the 22-string configuration (IceCube-22).

The expected neutrino flux from self-annihilation of a WIMP of mass $m_\chi$ is proportional 
to the square of the dark matter density, $\rho_{\mbox{\tiny DM}}^2$, integrated along the line of sight:
\begin{equation}
J_{\rm{a}}(\Psi) = \int\limits_0^{l_{\rm max}} \mathrm{d}l ~
	\frac{ \rho_{\mbox{\tiny DM}}^2 ( \sqrt{ R^2_{\mbox{\tiny SC}} - 2l R_{\mbox{\tiny SC}} \cos{\Psi} +l^2}) }
	{ R_{\mbox{\tiny SC}} \rho^2_{\mbox{\tiny SC}} }.
\label{eq:lineOfSightAnn}
\end{equation}
Here $\Psi$ is the angular distance of the line of sight 
with respect to the Galactic Center, and $R_{\mbox{\tiny SC}}$ and $\rho_{\mbox{\tiny SC}} = \rho(R_{\mbox{\tiny SC}})$ 
are used as scaling parameters to make $J_{\rm a}(\Psi)$ dimensionless. The upper limit of the integral is
\begin{equation}
	l_{\rm{max}} = \sqrt{R^2_{\mbox{\tiny MW}} - \sin^2{\Psi} R^2_{\mbox{\tiny SC}} } + R_{\mbox{\tiny SC}} \cos{\Psi},
	\label{eq:lineOfSightAnnLMax}
\end{equation}
where $R_{\mbox{\tiny MW}}$ is the radius of the Milky Way dark matter halo. Typically,
contributions from distances larger than the scale radius ($r_s$) of the halo model are
sufficiently small and can be neglected~\cite{Yuksel:2007ac}. A value of $R_{\mbox{\tiny MW}}=40$\,kpc
is adopted in this analysis.

\begin{figure}[t]
	\centering
	\includegraphics[width=0.5\textwidth]{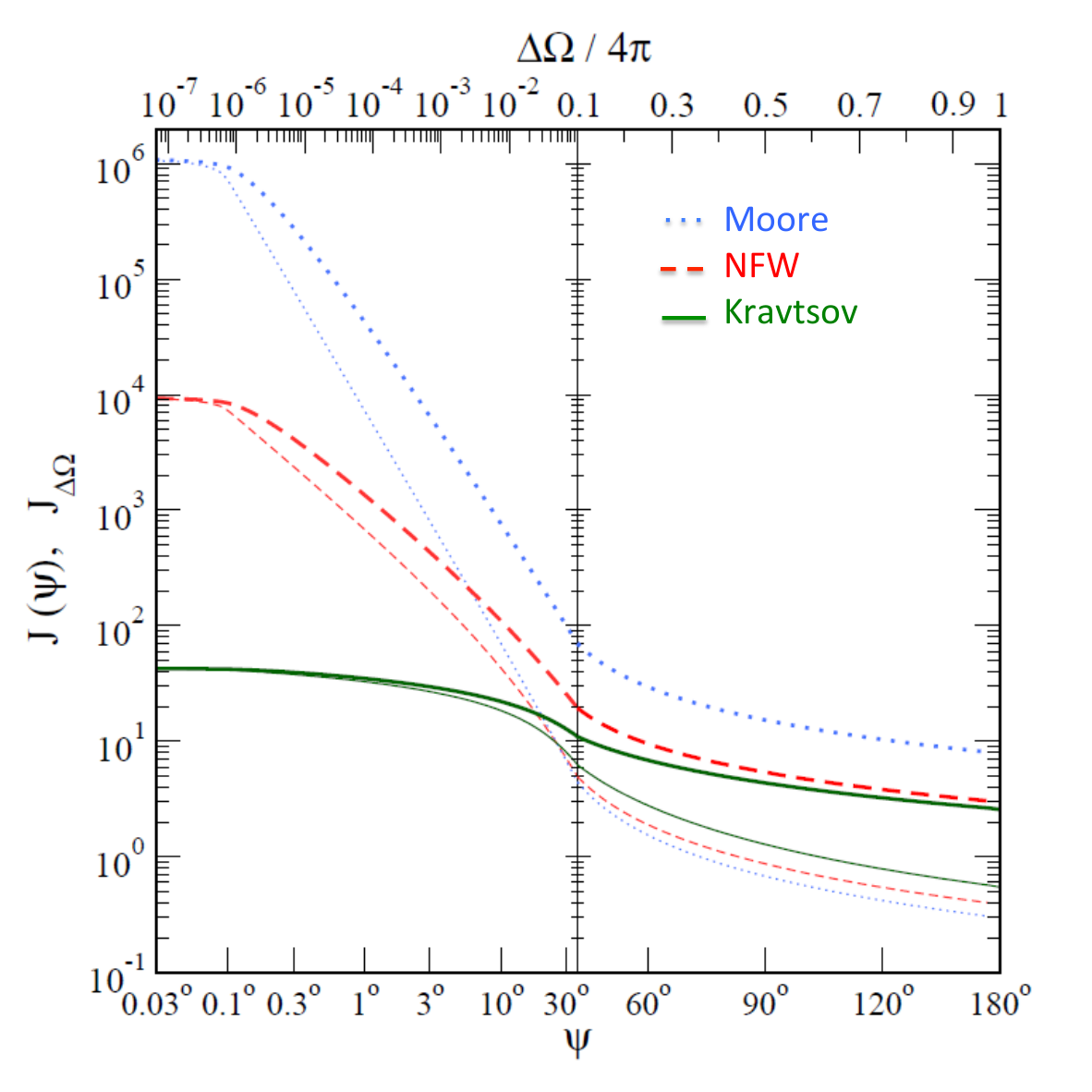}
	\caption{The line of sight integral, $J_{\rm a}(\Psi)$, and the average value of $J_{\rm a}$
		for the solid angle defined by the opening angle, $\Psi$, as a function
		of $\Psi$~\cite{Yuksel:2007ac}. The upper x-axis indicates the fraction
		of the solid angle defined by $\Psi$ with respect to the full sphere.
		The thin lines 	show the value of $J_{\rm a}$ at an opening angle of $\Psi$ while the thick
		lines depict $J_{\Delta \Omega }$ (which is the average value of $J_{\rm a}$) in the cone defined by $\Psi$.
		The green (solid) lines show the Kravtsov profile, the red (dashed) lines show the NFW and the blue (dotted)
		lines corresponds to the Moore profile.}
  \label{fig:JPsivsPsi}
\end{figure}

The quantity $J_{\Delta \Omega }$ is the average value of $J_{\rm a}(\Psi)$ for a chosen search region.
For a circular search region defined by the solid angle $\Delta \Omega$ it is described by
\begin{equation}
J_{\Delta \Omega } 
	%= \frac{\int_{\cos \Psi}^{1}J_{\rm a}(\Psi) \rm{d} (\cos \Psi')}{1- \cos\Psi}
	= \frac{2 \pi \cdot\int_{\cos \Psi}^{1} J_{\rm a}(\Psi') \rm{d} (\cos \Psi')}{\Delta \Omega}.
	\label{eqn:JOmega}
\end{equation}
Figure~\ref{fig:JPsivsPsi} shows $J_{\rm a}(\Psi)$ and $J_{\Delta \Omega }$ for the different
halo models and a circular search region.

In the above scenario the neutrino flux at Earth has two possible components; 
annihilation or decay of dark matter. The annihilation component may be written as~\cite{Yuksel:2007ac}
\begin{equation}
\frac{\rm{d}\phi_\nu}{\rm{d}E} =
	\frac{\left<\sigma_{\rm A} v\right>}
	{2} ~ J_{\rm a}(\Psi) ~ \frac{R_{\mbox{\tiny SC}}\rho_{\mbox{\tiny SC}}^2}{4\pi m_{\chi}^2} ~ \frac{\rm{d}\mathcal{N_{\nu}}}{\rm{d}E}.
\label{eq:galaxyfluxanni}
\end{equation}
The 1/(4$\pi$) factor accounts for the isotropic radiation of the annihilation products and
$\rm{d}\mathcal{N_{\nu}}/\rm{d}E$ gives the differential neutrino yield per annihilation.
The factors $\rho_{\mbox{\tiny SC}} /  m_{\chi} $ and $R_{\mbox{\tiny SC}}$ normalize
the dimensionless factor $J_{\rm a}$ to the number density, and the factor of $1\over 2$
accounts for the fact that two particles are required for annihilation.

The decay component is given by~\cite{PhysRevD.78.023502}
\begin{equation}
\frac{\rm{d}\phi_\nu}{\rm{d}E} = \frac{1}{\tau} ~ J_{\rm d}(\Psi) ~ \frac{R_{\mbox{\tiny SC}}\rho_{\mbox{\tiny SC}}}{4\pi m_{\chi}} ~ \frac{\rm{d}\mathcal{N_{\nu}}}{\rm{d}E}.
\label{eq:galaxyfluxdecay}
\end{equation}
Here $\tau$ is the lifetime of the dark matter particle and $J_{\rm d}(\Psi)$ is the
line of sight integral over the dark matter density,
\begin{equation}
J_{\rm d}(\Psi) = \int\limits_0^{l_{\rm max}} \mathrm{d}l ~ 
	\frac{ \rho_{\mbox{\tiny DM}} ( \sqrt{ R^2_{\mbox{\tiny SC}} - 2l R_{\mbox{\tiny SC}} \cos{\Psi} +l^2}) }
	{ R_{\mbox{\tiny SC}} \rho_{\mbox{\tiny SC}} }.
\end{equation}
\par

$\texttt{DARKSUSY}$~\cite{Gondolo:2004sc} is used to obtain the neutrino energy spectra
from WIMP-annihilations for various benchmark annihilation channels and WIMP masses.
Considered are specific masses of $100$\,GeV, $200$\,GeV, $ 300$\,GeV, $500$\,GeV,
$700$\,GeV, $1$\,TeV, $2$\,TeV, $5$\,TeV and $10$\,TeV,
and 100\% branching into $b\bar{b}$, $W^+W^-$, $\mu^+\mu^-$ or $\tau^+ \tau^-$ pairs.
The obtained energy spectra represent a model-independent conversion of the WIMP mass
to Standard Model particles and their subsequent decay. Therefore, the same spectra
can be used for the calculation of lifetimes in the case of decaying dark matter.

Annihilation and decay to $\nu\bar{\nu}$ produce line spectra at the WIMP mass
or half of the WIMP mass, respectively. The line spectrum is also interesting
from a theoretical point of view as it can be used to obtain model-independent and
conservative upper bounds for dark matter annihilation to Standard Model
final states~\cite{Beacom:2006tt}.

The choice of the annihilation channels cover a wide range of neutrino spectra.
The $b\bar{b}$ channel gives a softer energy spectrum while $\mu^+\mu^-$ and
$\nu~\bar{\nu}$ channels provide a harder energy spectrum. For a specific
supersymmetric scenario a superposition of these annihilation channels is
expected, producing an effective neutrino energy spectrum that lies between the
benchmark cases considered in this analysis. \par

The analysis presented here is based on a search for a muon-neutrino signal. For the calculation
of the muon-neutrino flux at Earth all neutrino flavors are considered assuming a
full mixing at Earth ($\nu_e:\nu_\mu:\nu_\tau$=1:1:1) due to long-baseline oscillations:
\begin{equation}
	\phi_{ \nu_{ \mu } } = \frac{1}{3}\sum_i \phi^0_{ \nu_i }.
\end{equation}
Here, $\phi^0_{ \nu_i }$ is the initial flux at source for each flavor.
Specific deviations from this benchmark assumption for vacuum oscillations, e.g.
as described in~\cite{PhysRevD.67.073024} ($\theta_{23}\approx \pi/4, \theta_{13}\approx 0, \sin^2{(2\theta_{12})}=0.86$),
lead to small changes in the predicted fluxes at the few percent level.

\begin{figure}[t]
  \centering
  \includegraphics[width=0.5\textwidth]{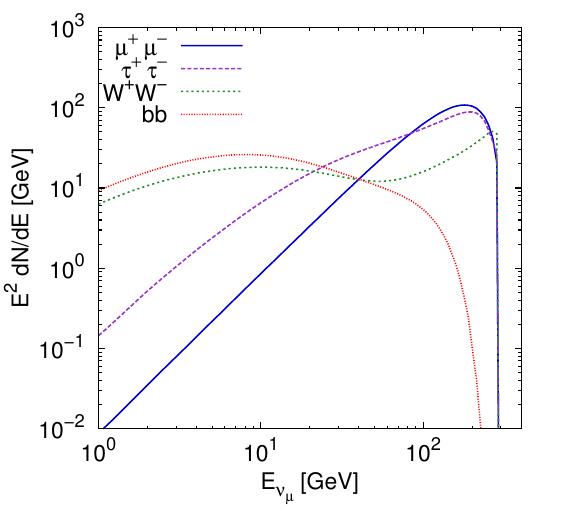}
  \caption{Energy spectra at Earth for different annihilation channels, including neutrino
	oscillations, for a 300\,GeV WIMP. The spectra were obtained with $\texttt{DARKSUSY}$~\cite{Gondolo:2004sc}.}
  \label{fig:signal_Edist}
\end{figure}

Figure~\ref{fig:signal_Edist} shows the expected energy distribution of
muon-neutrinos after full mixing at Earth for different annihilation
channels of a $300\,\mathrm{GeV}$ WIMP. Annihilation to $\nu\bar{\nu}$
produces a line spectrum at the WIMP mass. \par
 
The expected number of neutrino events is found by integrating
equations~(\ref{eq:galaxyfluxanni}) or~(\ref{eq:galaxyfluxdecay}) over
the detector live-time and the direction- and energy-dependent effective
area for annihilation or decay, respectively.
This procedure permits the direct conversion of an observed flux at the
detector to the velocity-averaged self-annihilation cross-section $\left<\sigma_{\rm A}
v\right>$, or the WIMP lifetime $\tau$.

\subsection{\label{sec:detector}The IceCube-40 Detector}

The IceCube in-ice neutrino detector is a cubic-kilometer array of digital optical
modules (DOMs) deployed between $1450\,\mathrm{m}$ and
$2450\,\mathrm{m}$ below the surface in the glacial ice of the South Pole. Each DOM
consists of a glass pressure housing enclosing a $25,4\,\mathrm{cm}$ diameter
Hamamatsu photomultiplier tube with digitizing electronics, and a set of LEDs
for calibration~\cite{Abbasi:2010vc, Abbasi:2008ym}. The DOMs are autonomous data
collection units with a time resolution of less than $3\,\mathrm{ns}$~\cite{Achterberg:2006md}.
They detect Cherenkov light emitted by the medium due to interaction with relativistic
charged particles, including charged leptons produced in neutrino interactions in the
ice, and muons produced by cosmic-ray interactions in the atmosphere. \par

The complete in-ice detector consists of 5160 DOMs deployed on 86 strings (60 DOMs per string).
Here 78 of the strings are arranged in a triangular pattern and form a nearly hexagonal
grid with 125\,m inter-string spacing and 17\,m vertical DOM separation.
The remaining 8 strings are located near the center of the detector interspersed with the
standard IceCube strings, and have an average inter-string separation of 72\,m and
vertical DOM spacing of 7\,m. Together with the surrounding 6 IceCube strings
and the central string, these strings form the low-energy sub-array called DeepCore~\cite{Abbasi2012615},
which is of particular interest for dark matter searches due to the lower energy threshold.

\begin{figure}[t]
  \centering
  \includegraphics[width=0.5\textwidth]{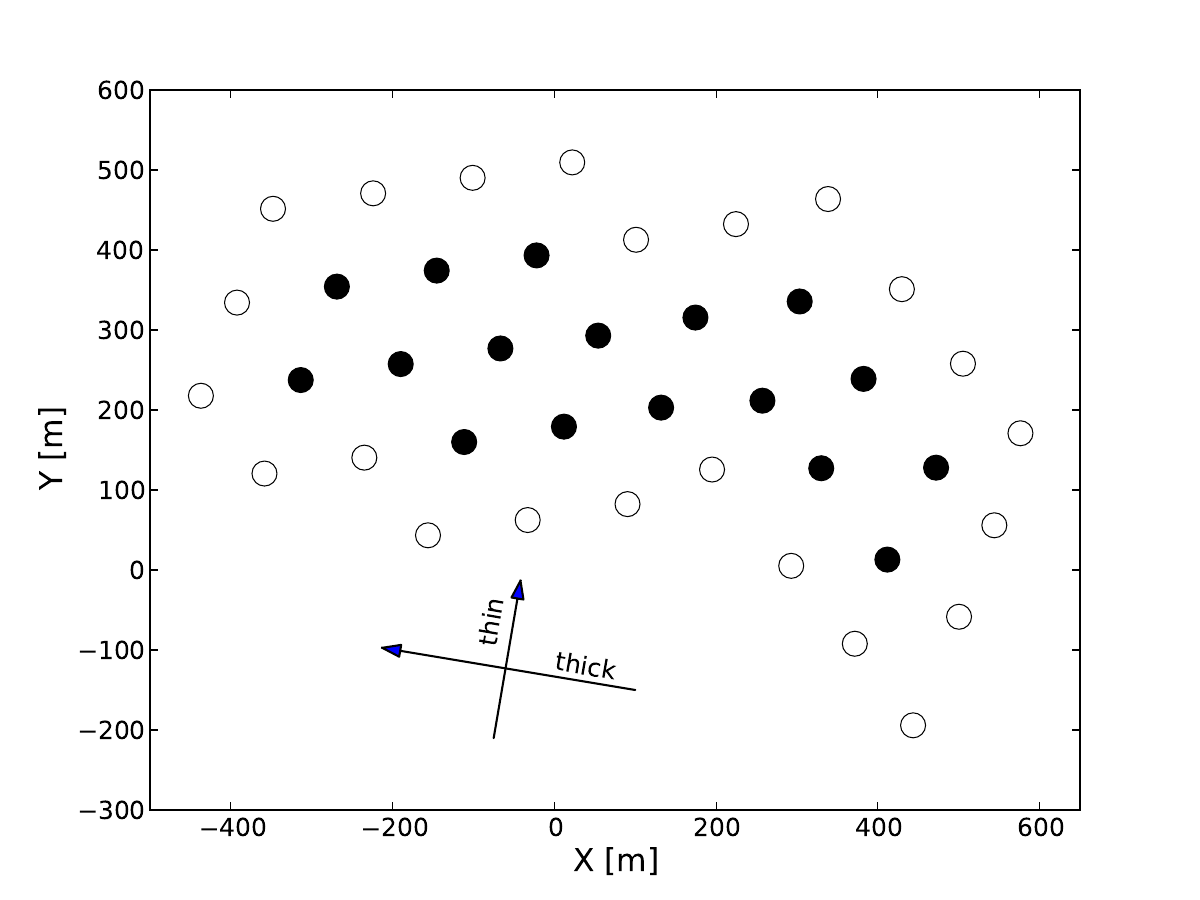}
  \caption{A sketch of string locations for the IceCube-40 detector (top view). The outer strings (open circles),
          which are used as a veto, as well as defined directions labeled as thick and
          thin are marked.}
  \label{fig:veto_geometry}
\end{figure}

The construction of IceCube was completed in December 2010, but the detector
was operational during the construction period from 2005 to 2010. In this paper we use data
from the 40-string configuration of IceCube (IceCube-40, see Fig.~\ref{fig:veto_geometry})
collected between April 2008 and May 2009, when IceCube did not contain any DeepCore strings.

The IceCube-40 data acquisition system applied several trigger algorithms
designed to collect different event topologies. This analysis considers events
that satisfy the SMT8 trigger, a simple multiplicity trigger, requiring at least
8 DOMs to record a signal above threshold within a time window of
$5\,\mu s$~\cite{Abbasi:2008ym}.

\subsection{\label{sec:analysis}Dataset and Analysis}

The data used represents 367 days of detector live-time~\cite{Huells2010:phd}.
Given IceCube's location at the South Pole, any search for a neutrino flux
from the direction of the Galactic Center, located in the southern sky
(RA $17^h$ $45^m$ $40.04^s$, Dec $-29^\circ$ 00$^\prime$ 28.1$^{\prime \prime}$), is
impacted by a significant background of cosmic-ray induced atmospheric muons
and atmospheric neutrinos.
The analysis strategy is to identify and select neutrino 
events which have an interaction vertex inside the detector
volume, since atmospheric muons, the dominant background, must enter from outside.
The large instrumented volume of IceCube-40 allows the definition of a veto region,
while retaining a sufficiently large fiducial volume for starting events.
The upper thirty DOMs on each string define a \emph{top-veto} while
the complete strings in the outer layer of IceCube-40 form a \emph{side-veto}
(Fig.~\ref{fig:veto_geometry}). Events with one or more hit in the top veto
or with the earliest hit recorded on any of the outer strings, are rejected.

Events surviving the veto condition are reconstructed with
an iterative log-likelihood track reconstruction method~\cite{Ahrens:04a}.
Based on the initial direction reconstruction, only events with a zenith angle larger
than 50$^\circ$ and an azimuth direction between 135$^\circ$ to 225$^\circ$
and 300$^\circ$ to 30$^\circ$, angular regions corresponding to the axis labeled as ``thick'' in Fig.~\ref{fig:veto_geometry},
are selected. For events outside these regions the detector is not thick enough to define an efficient veto.
Further cuts are applied to improve the angular resolution of the data.
Examples of such cuts include the reduced log-likelihood value from a track reconstruction
or the number of strings that have registered a hit in the recorded event.

To identify events with a neutrino interaction vertex within the detector
a likelihood algorithm has been developed that analyzes the event topology~\cite{Huells2010:phd}.
This algorithm determines the likelihood of a muon
to produce or not produce hits in the DOMs along the reconstructed track up-stream of
an assumed starting point and it includes the optical properties of the ice.
Maximizing the likelihood gives the location of the interaction vertex. The likelihood
ratio of a starting track relative to a through-going track hypothesis
is used as a classifier to further reduce the atmospheric muon background.

\begin{figure}[t]
	\centering
	\includegraphics[width=0.5\textwidth]{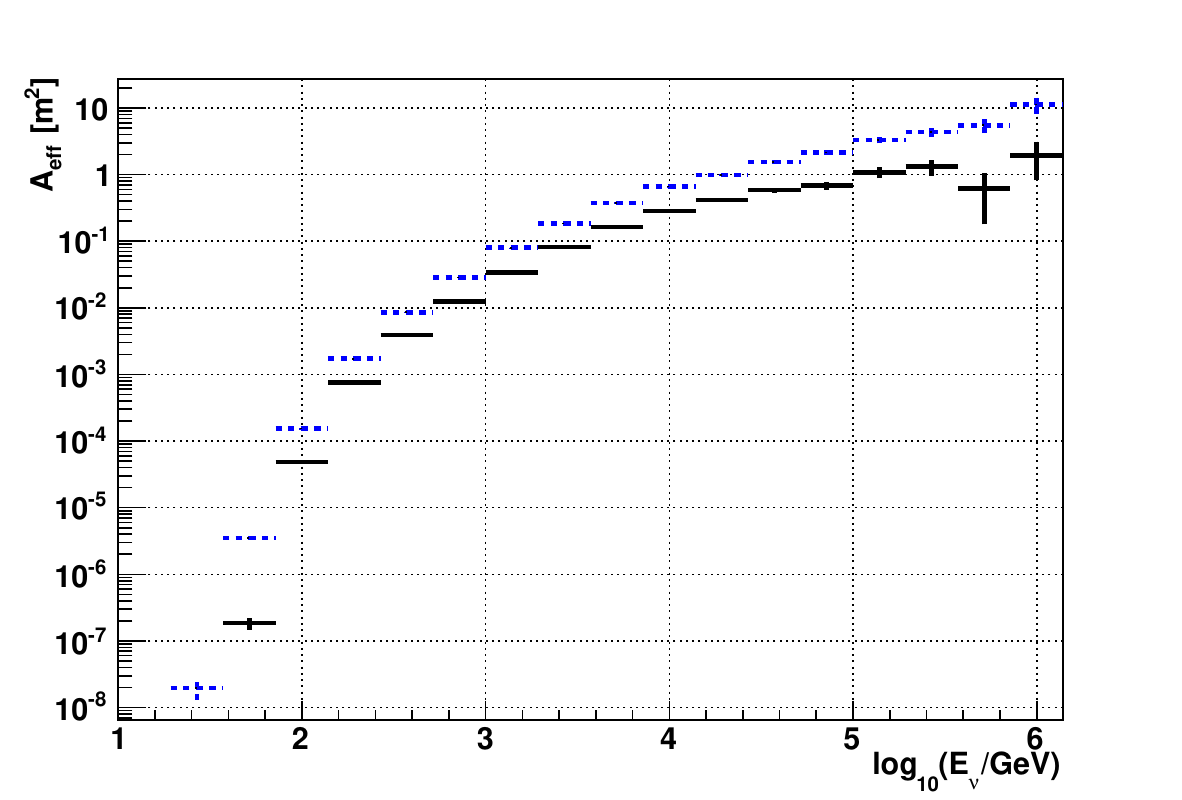}
	\caption{Effective area for neutrinos from MonteCarlo simulation for the direction of the
		Galactic Center after application of the veto conditions and angular cuts (blue-dashed)
		and after subsequent application of quality and signal selection cuts (black-solid).}
	\label{fig:a_eff}
\end{figure}

The above-mentioned quality and signal selection cuts reduce the dataset by about two
orders of magnitude. The event rate is reduced from an initial trigger rate of about 1\,kHz (SMT8) to 
2\,Hz (initial veto and azimuth cuts) to 0.56\,Hz (quality and signal selection cuts).
Figure~\ref{fig:a_eff} shows the effective area of the detector before and after the event
quality selection. The median angular resolution at this level is about 2.6$^\circ$.
The data remains dominated by cosmic-ray muons, undetected in the veto region,
that enter the detector from outside.\par
\begin{figure}[t]
	\centering
	\includegraphics[width=0.5\textwidth]{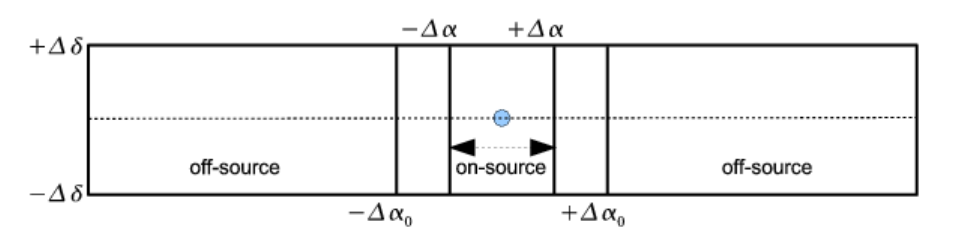}
	\caption{Definition of the on-source and off-source regions. The extent of the declination band is $2\Delta\delta$ for all regions.
		The width of the on-source region in right ascension is $2\Delta\alpha$. The region between
		$\Delta\alpha$ and $\Delta\alpha_{\mbox{\tiny 0}}$ is used as buffer. The region beyond
		$\Delta\alpha_{\mbox{\tiny 0}}$ is defined as off-source. In this analysis values of $\Delta\delta=8^\circ$,
		$\Delta\alpha = 9^\circ$, and $\Delta\alpha_{\mbox{\tiny 0}}=20^\circ$ are used.}
	\label{fig:on-off_region}
\end{figure}

This analysis uses an on-source/off-source method, as shown in Fig.~\ref{fig:on-off_region},
to estimate the background directly from the data. The highest flux of WIMP annihilation products is
expected to arrive from the direction of a wide region centered at the Galactic Center.
The atmospheric muon and neutrino background at the location of IceCube varies
with the zenith (or declination) angle but is approximately constant in right ascension.
Therefore, events from a declination band around the Galactic Center were selected. The width of the band 
in declination, $\pm \Delta \delta$, was optimized by maximizing $S/\sqrt{B}$,
where $S$ and $B$ are the number of expected signal and background events respectively 
for each WIMP mass and each annihilation channel.
Optimal bin sizes for $\Delta \delta$ were found to be between $6^\circ$ and $8^\circ$.
The variation of $S/\sqrt{B}$ with $\Delta \delta$ is rather small and a
value of $\Delta \delta = 8^\circ$ was adopted for all WIMP masses and annihilation channels.

The on-source region width in right ascension, $\Delta \alpha$, is chosen to be equal
to the size of the on-source region in declination at the zenith angle of the Galactic Center:
\begin{equation}
	\cos{\Delta \alpha} = \frac{\cos{(\Delta \delta)} - \cos^2{ (\theta_{\rm GC}) }}{\sin^2{ (\theta_{\rm GC}) }},
\end{equation}
where $\theta_{\rm GC}$ is the zenith angle of the Galactic Center in local coordinates
($61^\circ$). This yields a value for $\Delta \alpha \approx 9^\circ$.
No sharp transition between on- and off-source regions exists because the dark matter
density profile decreases continuously with distance from the Galactic Center. To reduce
the signal contamination of the background estimate from the off-source region, we define
a buffer between the on-source region and the off-source region (Fig.~\ref{fig:on-off_region}).
An event is considered off-source, and thus used for the background estimate, if the angular
distance to the Galactic Center is $\Delta \alpha_{\mbox{\tiny 0}} \geq 20^\circ$.
Events between $\Delta \alpha_{\mbox{\tiny 0}}$ and $\Delta \alpha$ are discarded.

The expected number of background events in the on-source region, $\non$, is:
\begin{equation}
	\non = \noff \cdot \frac{ \Delta \alpha }{ 180^\circ - \Delta \alpha_0}
	\label{eqn:onOffCalc}
\end{equation}
where $\noff$ is the number of background events in the off-source region.
This method of background estimation relies on measured data and is therefore
independent of Monte Carlo simulations and related systematic uncertainties
(e.g. modeling of optical ice properties or the detector response).\par

Despite the daily rotation of the detector variations of up to 1.5\,\% are found
in the right ascension distribution of the final event selection (Fig.~\ref{fig:expDist}).
The origin of these variations is the non-uniform azimuthal acceptance, caused by the asymmetric
geometrical layout of IceCube-40, combined with periods of detector inactivity which
are not evenly distributed in time. The arising uncertainty is larger than the expected
statistical uncertainty on the background estimate. A procedure has been applied to
account for the uneven exposure:

\begin{figure}[t]
  \centering
  \includegraphics[width=0.5\textwidth]{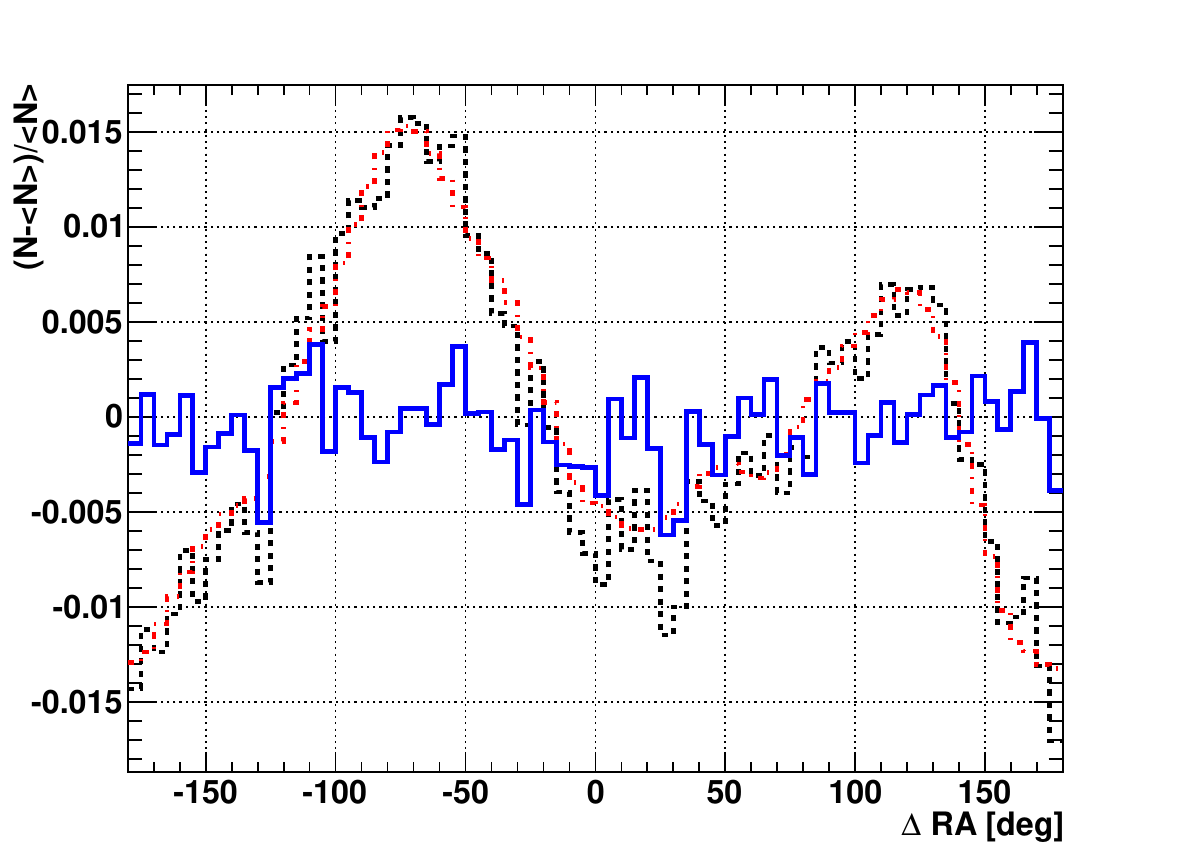}
  \caption{Relative variation of the event distribution in RA (with respect to the Galactic Center,
  thus the x-axis is labeled $\Delta RA$). $N$ is the number of measured events
  per bin, $\left < N \right >$ is the average number of events in all bins.
  The measured (black-dotted) and expected (red-dashed-dotted) distributions are shown,
  along with the deviation from expectation (blue).}
  \label{fig:expDist}
\end{figure}

\begin{figure}[t]
  \centering
  \includegraphics[width=0.5\textwidth]{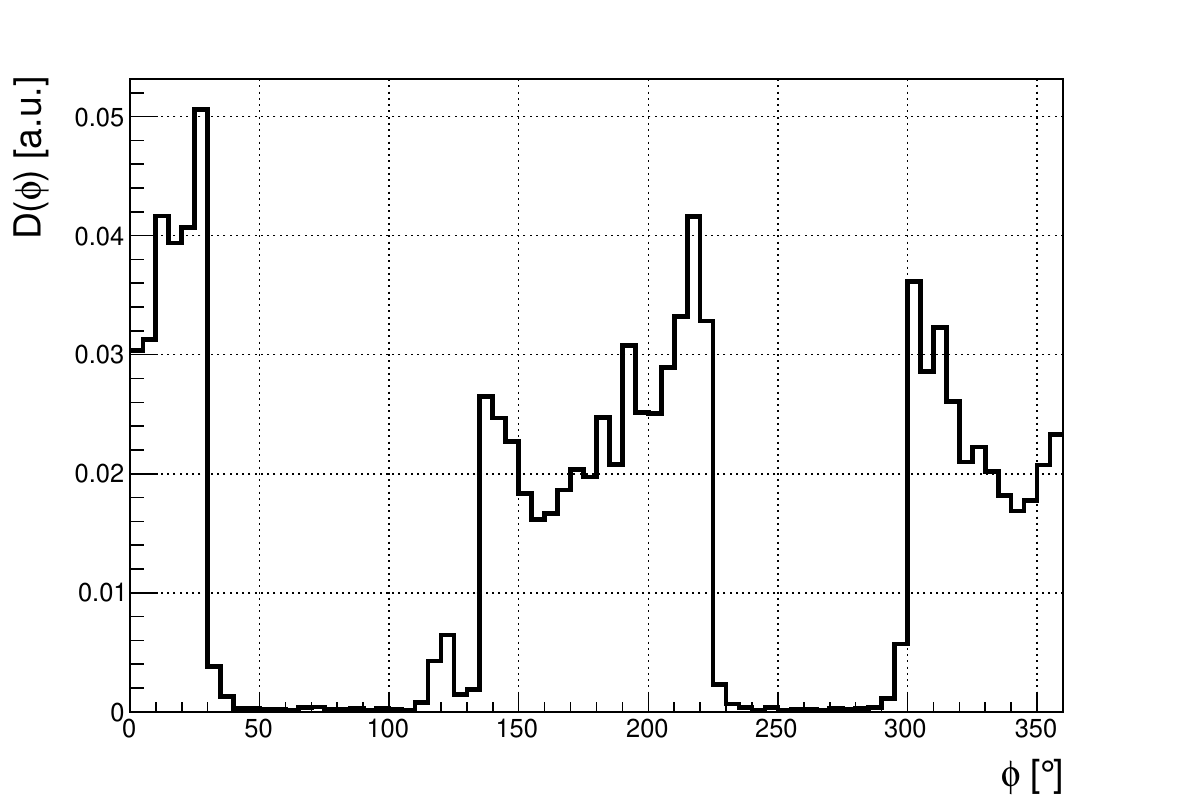}
  \caption{Initial azimuth pdf containing all events from the off-source region. The cuts on
           the reconstructed azimuth were performed on an initial reconstruction. This figure shows the
           	azimuth distribution based on an advanced reconstruction. The cut regions excluding events
           	from the ``thin'' (Fig.~\ref{fig:veto_geometry}) detector orientation are clearly visible.}
  \label{fig:azimuthPDF}
\end{figure}

\begin{itemize}
	\item First, we determine the normalized distribution, $D(\phi)$, of
				all measured event azimuth angles, $\phi$, in detector coordinates that do not originate
				in the on-source region. This distribution, shown in Fig.~\ref{fig:azimuthPDF},
				corresponds to the effective probability density function (pdf) for the
				observation of an event from a certain detector direction, integrated over the
				exposure in time. This pdf rotates with the detector and is not the exposure in
				right ascension.
	\item	Second, we transform for each event time the full pdf to equatorial
				coordinates (RA,$\delta$): $D'_i(RA) = D(RA + \Delta \phi_i)$. At IceCube's location,
				this corresponds to a rotation of the azimuth pdf by an angle, $\Delta \phi_i$,
				that can be calculated from the event time.
	\item Lastly, we calculate the sum of all the obtained distributions (in RA):
				\begin{equation}
					D'(RA) = \sum\limits_i^N D'_i = \sum\limits_i^N D(RA + \Delta \phi_i),
				\end{equation}
				where $N$ is the total number of events.
\end{itemize}
The application of this procedure results in an exposure-corrected expected distribution
of background events in right ascension, normalized to the number of events in the full dataset.
Figure~\ref{fig:expDist} shows the accuracy of the number of predicted events.
The agreement between the number of measured and expected events is at the 0.1\,\%-level.
This approach also takes seasonal variations into account. \par

\subsection{\label{sec:systematics}Systematic Uncertainties}

Predictions of the expected signal are dominated by large variations in the
description of the central region of halo models (Fig.~\ref{fig:HaloModels}).
In particular, in the direction of the Galactic Center, the line of sight integral, $J(\Psi ) $,
spans several orders of magnitude (Fig.~\ref{fig:JPsivsPsi}). For this reason we have
chosen the NFW profile as a benchmark model, and do not consider the halo-model
dependence a systematic uncertainty.

The limits obtained in this analysis can be rescaled to limits for the
Moore or Kravtsov profile by multiplication of the limit for each mass
or channel with $J^{NFW}_{\Delta \Omega }/J'_{\Delta \Omega }$, where
$J'_{\Delta \Omega }$ is the desired profile. The corresponding values
of $J_{\Delta \Omega }$ were obtained for the actual box-shaped search bin
(not by means of equation~(\ref{eqn:JOmega})) and are shown in Table~\ref{tab:JVals}. 
However, especially for flat-cored profiles the signal-contamination of the
off-source region has to be considered when rescaling the limits.

\begin{table}
	\caption{$J_{\Delta \Omega }$ for annihilating dark matter for the considered halo
	density profiles and the chosen on-source region.}
	\label{tab:JVals}
	\begin{tabular}{l|c|c|c}
		\hline
		\hline
			 & NFW & Kravtsov & Moore \\
		\hline
		$J_{\Delta \Omega } $& 119 & 15	& 1209 \\
		\hline
		\hline
	\end{tabular}
\end{table}

The background prediction is derived from the high-statistics off-source data sample.
The systematic uncertainty in this prediction is limited to small directional
variations in the cosmic ray flux, e.g. as reported in~\cite{0004-637X-740-1-16}.
A Kolmogorov-Smirnov test for compatibility of the relative deviations to a
normal distribution yields a p-value of 0.77. Within the limited statistics of 72
values ($5^\circ$-bins), the result does not hint at a possible anisotropy in this data sample.

Though the background estimate is data-driven, systematic uncertainties arise from
the Monte Carlo-based estimation of the detector sensitivity for signal events.
Typically, the energies considered in this analysis are lower than  
for other IceCube analyses, e.g.\ the search for high energy extragalactic neutrinos,
and are largely similar to the energies considered in Ref.~\cite{Abbasi:2011eq,PhysRevD.85.042002}.
For lower energies, the uncertainties on the neutrino cross-section and
neutrino and muon propagation are on the order of $7$\,\%~\cite{PhysRevD.85.042002}. 
In addition, uncertainties corresponding to the detector up-time are less than 1\%.
Uncertainties related to the detection efficiency of low energy events, such as the
sensitivity of optical modules~\cite{Abbasi:2010vc} and the photon propagation through
the ice~\cite{IceCube:icepaper}, are considerably larger. These effects lead to a typical uncertainty 
on the total effective area of about 25\,\% for lower energies, $\mathcal{O}(100\,\mathrm{GeV})$,
and about 20\,\% for energies near $1$\,TeV.
Though the impact of the above-mentioned systematics is energy-dependent
we conservatively assume a 33\,\% effect on the signal detection efficiency, $\epsilon_{\rm det}$.
Here, $\epsilon_{\rm det}$ corresponds to the integration over energy of the neutrino multiplicity
per annihilation multiplied by the effective area:
\begin{equation}
	\epsilon_{\rm det} \propto \int \mathrm{d} E A_{\rm eff}(E) \frac{\rm{d}\mathcal{N_{\nu}}}{\mathrm{d}E}.
\end{equation}
The value of 33\,\% is propagated into the limits by means of equation~(\ref{eq:galaxyfluxanni}),
\begin{equation}
	\left<\sigma_{\mathrm A} v\right> \propto  \frac{N_{ \mathrm{lim} } }{\epsilon_{\mathrm{det}} \cdot (1 \pm 0.33)},
\end{equation}
for all channels and masses, where $N_{\rm lim}$ is the limit on the number
of signal neutrinos in the sample. The systematic uncertainty on the lifetimes
have been included by means of equation~(\ref{eq:galaxyfluxdecay}),
\begin{equation}
	\tau \propto  \frac{\epsilon_{\mathrm{det}} \cdot (1 \pm 0.33)}{N_{ \mathrm{lim} } }.
\end{equation}
The presented limits have been shifted upwards by a factor of 1/0.67 to include
systematic effects in a conservative way.

\begin{table}
	\caption{Summary of systematic effects.}
	\label{tab:sys}
	\begin{tabular}{l|c}
		\hline
    	\hline
		Effect	&	Contribution \\
		\hline
		live-time	& $<1$\,\% \\
		\hline
		$\nu$ cross-section and \\ propagation  & 7\,\%  \\
		\hline
		photon propagation and \\ optical ice properties & 20\,\%-25\,\% \\
		\hline
		total	(conservative)	&  33\,\% \\
		\hline
		\hline				
	\end{tabular}
\end{table}

\subsection{\label{sec:results}Results}

\begin{figure}[t]
	\centering
	\includegraphics[width=0.45\textwidth]{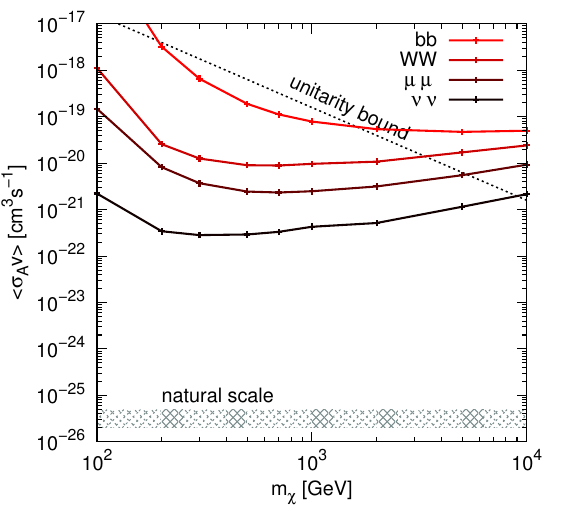}
	\caption{ Upper limits (90\,\% C.L.) on the self-annihilation cross-section for the four channels at
  	nine WIMP masses.
  	The hashed area represents the natural scale at which WIMPs are thermal relics~\cite{Steigman:2012nb}.
  	The black-dotted line is the unitarity bound~\cite{Griest:1989wd, Hui:2001wy}.
  	The four limits were derived assuming the NFW density profile.}
	\label{fig:IceCube40Limits}
\end{figure}

\begin{figure}[t]
	\centering
	\includegraphics[width=0.45\textwidth]{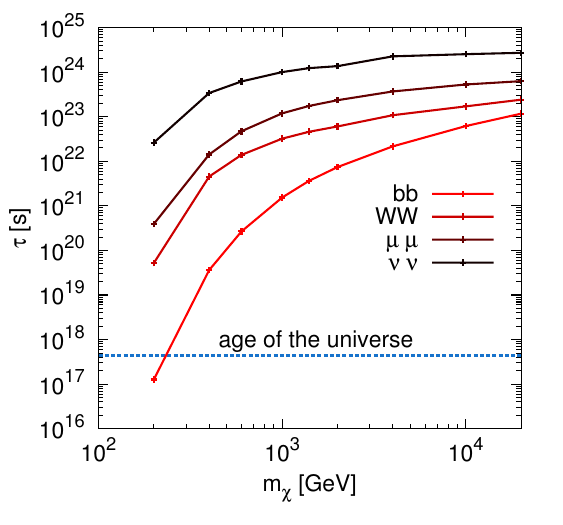}
	\caption{Lower limits (90\,\% C.L.) on the lifetime for the four channels at nine WIMP masses.
  				The horizontal line represents the age of the universe~\cite{Jarosik:2010iu}.
  				The four limits were derived assuming the NFW density profile.}
  \label{fig:IceCube40LimitsDecay}
\end{figure}

The number of observed events in the on-source region was
$N^{ \mbox{\rm \tiny{on}} } = 798\,842$, while $\non = 798\,819$
background events were expected based on estimates from the approximately 5 times larger
larger off-source region, using equation~(\ref{eqn:onOffCalc}). This procedure yields a
relatively small uncertainty on the background expectation.
If the null hypothesis is valid, the outcome of a large number of measurements
of $N^{ \mbox{\rm \tiny{on}} }$ is expected to follow a Poisson distribution around $\non$.
The small difference of $\Delta N = N^{ \mbox{\rm \tiny{on}} } - \non = 23$ events,
compared to a standard deviation of about 916 events, is compatible with no neutrino
signal from annihilations of dark matter particles.
Based on Feldman-Cousins~\cite{PhysRevD.57.3873}, an upper limit (90\,\% C.L.)
of 1504 signal neutrinos is computed.
From equation~(\ref{eq:galaxyfluxanni}), the limit on the velocity-averaged WIMP
self-annihilation cross-section $\left<\sigma_{\rm A} v\right>$ for annihilations in
the Galactic Center is derived for various WIMP masses and annihilation
channels (Fig.~\ref{fig:IceCube40Limits}).
Similarly, with equation~(\ref{eq:galaxyfluxdecay}), lower limits on the WIMP
lifetime are calculated (Fig.~\ref{fig:IceCube40LimitsDecay}).

\subsection{\label{sec:fermipamela}Comparison with Results from Other Indirect Searches}

\begin{figure}[t]
  \centering
  \includegraphics[width=0.5\textwidth]{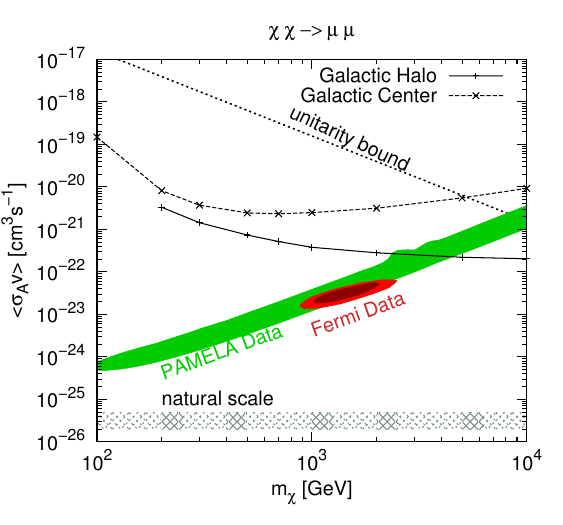}
  \caption{ Upper limits (90\,\% C.L.) on the self-annihilation cross-section for the annihilation channel to muons at nine WIMP masses assuming the NFW density profile (solid line). The hashed area represents the natural scale at which WIMPs are thermal relics~\cite{Steigman:2012nb}. The black-dotted line is the unitarity bound~\cite{Griest:1989wd, Hui:2001wy}. The green ($3\sigma$) region is based on the PAMELA excess, the red region represents the PAMELA data combined with the Fermi and H.E.S.S data ($3\sigma$ and $5\sigma$)~\cite{Meade:2009iu}.}
  \label{fig:IceCube40LimitsComp}
\end{figure}

Several satellite-borne and balloon experiments have reported deviations of
measured cosmic-lepton spectra from theoretical predictions. The PAMELA
collaboration observed an increased positron fraction at energies above
$\sim$\,10\,GeV~\cite{Adriani:2008zr}, that disagrees with the
conventional model for cosmic-rays~\cite{Moskalenko:1997gh} (secondary positron production only).
This measurement has also been confirmed by the Fermi-LAT collaboration~\cite{PhysRevLett.108.011103}.
Further, the Fermi-LAT~\cite{Abdo:2009zk} and HESS~\cite{Aharonian:2009ah,PhysRevLett.101.261104}
measurements of primary electrons showed a discrepancy when compared to conventional diffuse models.
Though the results may be interpreted as signals from nearby astrophysical
sources~\cite{Yuksel:2008rf}, they could also be explained by annihilation or
decay of leptophilic dark matter with a mass of $\mathcal{O}$(1\,TeV).
With this assumption, preferred regions in the \sacs-$m_\chi$-plane
can be identified~\cite{Meade:2009iu, Cirelli:2008pk} and compared to
the limits obtained in this analysis. Figure~\ref{fig:IceCube40LimitsComp}
shows our $\mu^+ \mu^-$ channel limits compared to those regions.

\begin{figure}[thb]
  \centering
   \includegraphics[width=0.5\textwidth]{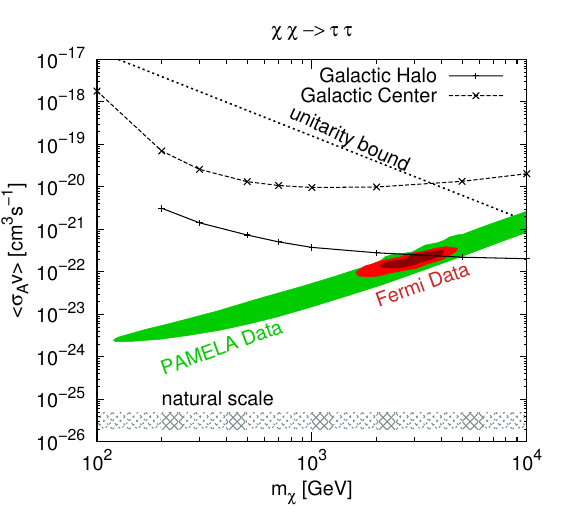}
   \caption{ Upper limits (90\,\% C.L.) on the self-annihilation cross-section for the annihilation channel to tau leptons at nine WIMP masses assuming the NFW density profile (solid line). The hashed area represents the natural scale at which WIMPs are thermal relics~\cite{Steigman:2012nb}. The black-dotted line is the unitarity bound~\cite{Griest:1989wd, Hui:2001wy}. The green ($3\sigma$) region is based on the PAMELA excess, the red region represents the PAMELA data combined with the Fermi and H.E.S.S data ($3\sigma$ and $5\sigma$)~\cite{Meade:2009iu}.}
  \label{fig:IceCube40LimitsCompTau}
\end{figure}

Figure~\ref{fig:IceCube40LimitsCompTau} shows a similar comparison for the
$\tau^+ \tau^-$ channel
% where the current limits are slightly closer to the
%preferred regions of the combined PAMELA and Fermi data. The better exclusion
%power in the $\tau$-channel is not due to a more stringent limit, but rather the
%location of the favored regions since there is significant energy loss during electron
%propagation, any electron or positron signal must originate from nearby sources
%($\mathcal{O}$(1\,kpc)). Given this, the shape of the preferred regions does not
%depend strongly on the halo profile.

\begin{figure}[t]
  \centering
  \includegraphics[width=0.45\textwidth]{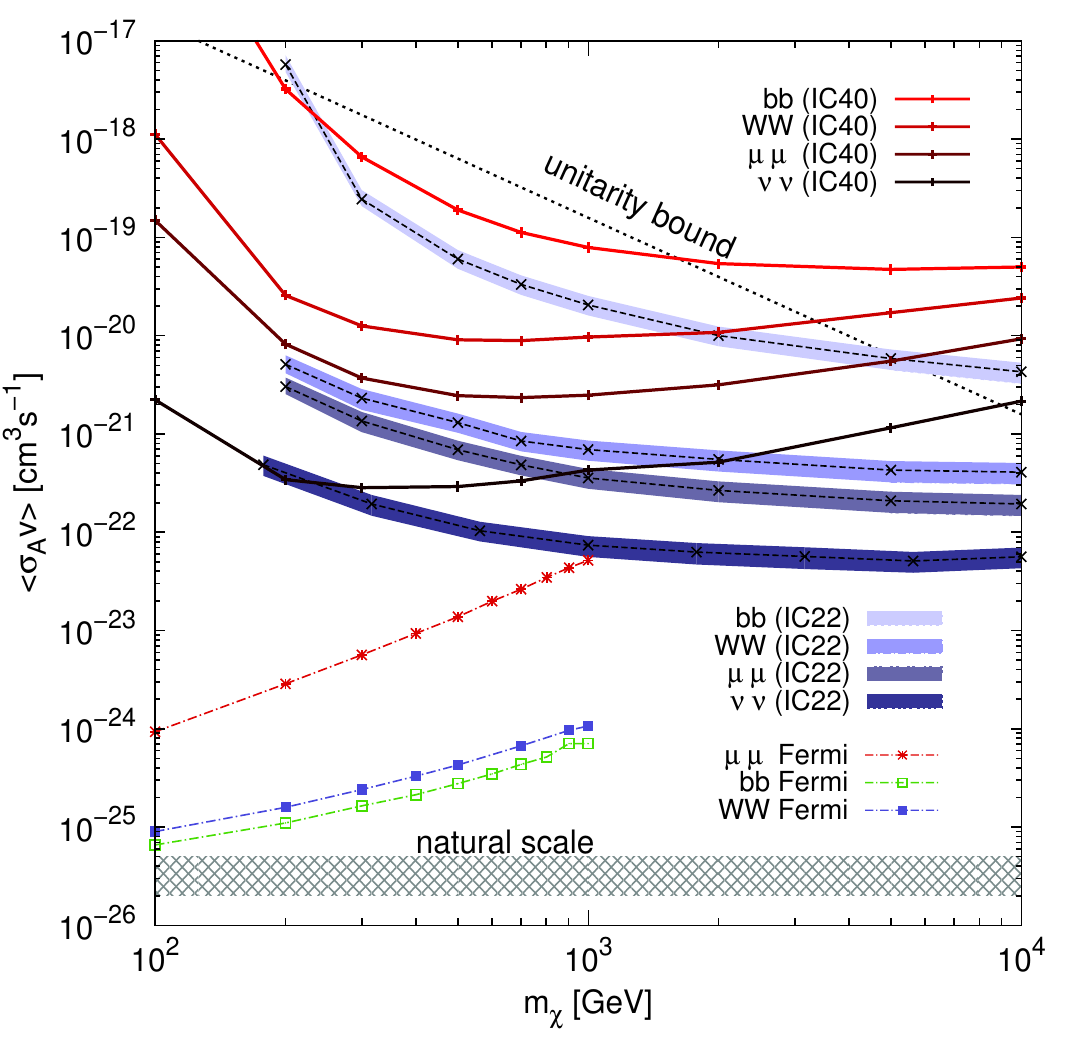}
  \caption{The upper limits from this analysis compared to those from an analysis of the
  outer Galactic halo using IceCube-22 data~\cite{Abbasi:2011eq}, as well as limits
  obtained from observation of dwarf galaxies by the Fermi-LAT collaboration~\cite{PhysRevLett.107.241302}.}
  \label{fig:IC22comp}
\end{figure}

Recently, the  Fermi-LAT collaboration~\cite{PhysRevLett.107.241302}
reported results from a search for dark matter annihilation signals
from 10 dwarf galaxies. For gamma-ray experiments dwarf galaxies are
very promising targets due to a relatively high signal-to-noise ratio.
The obtained limits are very stringent and a comparison to our Galactic Center
limits, as well as the IceCube-22 Galactic halo limits~\cite{Abbasi:2011eq},
is shown in Fig.~\ref{fig:IC22comp}. At lower WIMP masses the dwarf limits
are more constraining than the IceCube limits by several orders of magnitude.
However, the IceCube limits on direct annihilation to neutrinos can be
regarded as a very conservative bound on the total dark matter annihilation
to Standard Model final states~\cite{Beacom:2006tt}. At higher WIMP masses ($\ge \mathrm{TeV}$),
where there is larger neutrino cross-section and muon range, IceCube benefits
from the large instrumented volume due to the increasing neutrino cross-section
and the muon range~\cite{IceCube:2011ae}.

In a different approach, the dark matter self-annihilation cross-section can also be
constrained based on precise measurement of the cosmic microwave background (CMB).
A too large self-annihilation cross-section can cause significant distortions of the
CMB depending on the annihilation mode. The lack of such distortions can be translated to
a limit on \sacs.  Very stringent limits were also obtained~\cite{PhysRevD.80.023505,PhysRevD.84.027302}
from measurements of temperature, anisotropy and polarization of
the CMB, based on WMAP~\cite{Jarosik:2010iu} and ACT~\cite{0004-637X-722-2-1148} data. However 
we refrain from comparing the derived limits to our results here, as our analysis targets
dark matter properties in the present epoch, while conditions in the early  universe were 
substantially different, e.g. the velocity distribution, as well as the energy absorption processes
in the photon-baryon plasma~\cite{PhysRevD.80.043526}.\par


\subsection{\label{sec:outlook}Summary and Outlook}

 The first search for dark matter accumulated in the center of the galaxy with neutrinos 
has been presented, using the 40-string configuration of IceCube. The analysis uses a fiducial 
detector volume surrounded by a layer of veto strings in order to define starting tracks and 
allow for efficient rejection of downgoing atmospheric muons. Limits were obtained for a wide 
range of WIMP masses and annihilations channels, and are compared in Figure~\ref{fig:IC22comp} 
with previous IceCube limits from an  analysis of the outer halo with IceCube-22~\cite{Abbasi:2011eq}. 
The analysis was able to extend the limits to lower WIMP masses (100~GeV) with respect to the 
IceCube-22 limits, showing the potential benefits of the veto technique. 

 However the sensitivity of this analysis is limited by the significant atmospheric muon background 
from the direction of the Galactic Center as well as the relatively high detection threshold. The need for a strong 
backgroud reduction imposes the need for a large top-veto region, and results in a considerable reduction 
in the fiducial volume of the detector. Besides, a one-side veto layer is still not optimal for the 
complete rejection of downgoing atmospheric muons reaching the detector laterally. The results obtained 
are therefore not fully competitive with the IceCube-22 results from the Galactic Halo, which did 
not need such a strong background rejection.

 Substantial improvement is expected from analyses using the considerably larger detector configuration with 
79 strings, IceCube-79, and the complete IceCube-86 detector. A larger detector provides a larger fiducial 
volume while maintaining a sufficiently large and effective veto volume to suppress the atmospheric
background~\cite{startingtracks}. In addition, preliminary estimates suggest that the DeepCore
low-energy extension, first present in IceCube-79, increases the effective area significantly
in the energy region between 10\,GeV and 100\,GeV.
Dedicated data filters for the Galactic Center have been developed and implemented for
IceCube-79~\cite{IceCube:2011ae}, and further improved in terms of signal acceptance
for the complete IceCube-86 detector.

%\nolinenumbers
\subsection*{\label{sec:ack}Acknowledgments}

We acknowledge the support from the following agencies:
U.S. National Science Foundation-Office of Polar Programs,
U.S. National Science Foundation-Physics Division,
University of Wisconsin Alumni Research Foundation,
the Grid Laboratory Of Wisconsin (GLOW) grid infrastructure at the University of Wisconsin - Madison, the Open Science Grid (OSG) grid infrastructure;
U.S. Department of Energy, and National Energy Research Scientific Computing Center,
the Louisiana Optical Network Initiative (LONI) grid computing resources;
National Science and Engineering Research Council of Canada;
Swedish Research Council,
Swedish Polar Research Secretariat,
Swedish National Infrastructure for Computing (SNIC),
and Knut and Alice Wallenberg Foundation, Sweden;
German Ministry for Education and Research (BMBF),
Deutsche Forschungsgemeinschaft (DFG),
Research Department of Plasmas with Complex Interactions (Bochum), Germany;
Fund for Scientific Research (FNRS-FWO),
FWO Odysseus programme,
Flanders Institute to encourage scientific and technological research in industry (IWT),
Belgian Federal Science Policy Office (Belspo);
University of Oxford, United Kingdom;
Marsden Fund, New Zealand;
Australian Research Council;
Japan Society for Promotion of Science (JSPS);
the Swiss National Science Foundation (SNSF), Switzerland.
We are grateful to Michael Gustafsson for the communication on the lifetime limit.

\bibliography{}
\hyphenation{Post-Script Sprin-ger}

\end{document}